\documentclass{kluwer}
\begin{document}
\begin{article}
\begin{opening}
\title{\bf
BLACK HOLE BINARY DYNAMICS
}
\author{
SVERRE J \surname{AARSETH}
}
\institute{Institute of Astronomy, Madingley Road, Cambridge CB3 0HA \\
E-mail: sverre@ast.cam.ac.uk}

\begin{abstract}
We discuss a new $N$-body simulation method for studying black hole
binary dynamics. This method avoids previous numerical problems due to
large mass ratios and trapped orbits with short periods.
A treatment of relativistic effects is included when the associated
time-scale becomes small.
Preliminary results with up to $N = 2.4 \times 10^5$ particles are
obtained showing systematic eccentricity growth until the relativistic
regime is reached, with subsequent coalescence in some cases.
\end{abstract}
\end{opening}

\def\ZZ#1{$\scriptstyle #1$}

\section{Introduction}

The problem of the formation and dynamical evolution of a black hole (BH)
binary with massive components is of considerable topical interest.
Several previous efforts employed direct integration methods to elucidate
the behaviour of such systems but applications to galactic nuclei pose
severe limitations with regard to the particle number which can be
investigated.
The formation is usually envisaged as the end product of two separate
galactic nuclei (or one being a dwarf galaxy) spiralling together by
dynamical friction, but there are other scenarios.

Notwithstanding the particle number limitations imposed by the need for
accurate numerical treatment, much can be learnt about the early evolution
of BH binaries by studying smaller systems.
It is well established that the presence of two massive bodies in a
stellar system leads to their rapid inward spiralling and inevitable
formation of a dominant binary.
In fact, this development was already discussed 30 years ago
(Aarseth 1972) for a relatively small system of $N = 250$ particles.
Here a massive binary absorbed about 90\% of the total cluster energy
after only 50 initial dynamical (or crossing) times.
This calculation was also the earliest demonstration of two-body
regularization in an $N$-body context.

Upon formation, the subsequent binary evolution is subject to a steady
shrinkage of the semi-major axis and ejection of particles resulting from
sling-shot interactions.
Although the rate of shrinkage decreases significantly, the corresponding
energy increase is fairly constant (Quinlan \& Hernquist 1997,
Milosavljevi\'c \& Merritt 2001).
These investigations employed two-body regularization in order to reduce
the systematic errors associated with direct integration of hard binaries.
Alternative studies based on a small softening of the Newtonian potential
also yielded similar results (Makino 1997).
The long-term evolution is characterized by significant depletion of the
central region which ceases to be representative of a realistic system.
However, even the use of chain regularization (Mikkola \& Aarseth 1993) for
treating compact subsystems is not sufficient to prevent numerical problems.

\section{Special Binary Treatment}

In view of the numerical problems outlined above, it is highly desirable
to develop a more suitable integration method.
A critical appraisal of chain regularization with two massive members
reveals that their contribution to the Hamiltonian energy dominates
and hence solutions of the equations of motion for any other members are
subject to the loss of precision.
This recognition led to the construction of a new method which is based
on kinematical considerations (Mikkola \& Aarseth 2002).
Briefly, a special time transformation is combined with the standard
leapfrog scheme, thereby avoiding a Hamiltonian formulation.
This allows extremely close two-body encounters to be studied without
significant loss of accuracy.
The interested reader is referred to the published description for
more details.

The new method is based on including all the $N (N-1)/2$ interaction
terms and solving the equations of motion by the Bulirsch--Stoer (1966)
integrator.
However, all the solutions need to be advanced with the same time-step
which limits the practical membership severely.
Hence this formulation can only be used to describe the motions of
a compact subsystem.
The implementation of such a solution method into a large $N$-body
simulation code is somewhat analogous to that for chain regularization
and has been outlined elsewhere (Mikkola \& Aarseth 1993, Aarseth 1999).
In the following we comment on some special features of the BH scheme.

The vital question concerning binary BH evolution is whether a stage
can be reached where the gravitational radiation time-scale is
sufficiently short for coalescence or significant shrinkage to occur.
Previous simulations did not address this issue, mainly because the
calculations were terminated prematurely for technical reasons.
In the present scheme we have included the 2.5 post-Newtonian
approximation for the most critical two-body interaction (Soffel 1989).
An estimate of the smallest semi-major axis which can be reached in a
system of $N$ particles can readily be made for a given mass ratio.
This size is several orders of magnitude outside the relevant range
for reasonable system parameters.
However, small two-body separations can also be achieved if the
eccentricity becomes large enough.

Although large eccentricities were not reported by other investigators,
more careful calculations do show significant eccentricity growth during
the late stages.
Hence this behaviour justifies the extra cost of including the
relativistic terms, but only when the corresponding time-scale is less
than the expected calculation time.
We have $a/\dot a \propto a^4 (1 - e^2)^{7/2}$ for the decay time,
where $a$ is the semi-major axis and $e$ is the eccentricity.
Since the time-scale is quite long for circular orbits, we need large
values of $e$ before activating the relativistic treatment which is
implemented at different levels of complexity.
Thus we distinguish between the classical radiation term and two
different expansion orders which describe the post-Newtonian acceleration
and relativistic precession, respectively.
Finally, coalescence is defined to take place if the BH separation
becomes less than three Schwarzschild radii.

\section{Numerical Results}

The initial conditions consist or two cuspy dwarf galaxy models with $N_0$
equal-mass particles of mass $\bar m$ at the apocentre of an eccentric
orbit ($e=0.8$) having a separation of $8 r_{\rm h}$, where $r_{\rm h}$
is the local half-mass radius.
A single BH of mass $m_{\rm BH} = (2 N_0)^{1/2} \bar m$ is placed at the
centre of each system.
The availability of the special-purpose GRAPE-6 supercomputer together with
a fast workstation host allows quite large particle numbers to be studied.
Here we report briefly on two recent simulations with $N_0 = 6 \times 10^4$
and $1.2 \times 10^5$ particles, making a total of $1.2 \times 10^5$ and
$2.4 \times 10^5$ members, respectively.
The two mass distributions soon combine into one slightly elongated
system, with the dominant binary already formed at the centre after only
about 20 crossing times in both models.
Then follows a period of constant energy gain where the BH binary is
advanced by standard two-body regularization.
A switch is made to the new method when the binary becomes super-hard;
i.e. $a \le 10^{-4} r_{\rm h}$.
The subsequent slow evolution necessitates a large number of perturbed
binary orbits to be studied.

\begin{figure}[t]
\vspace{8.5cm}
\includegraphics{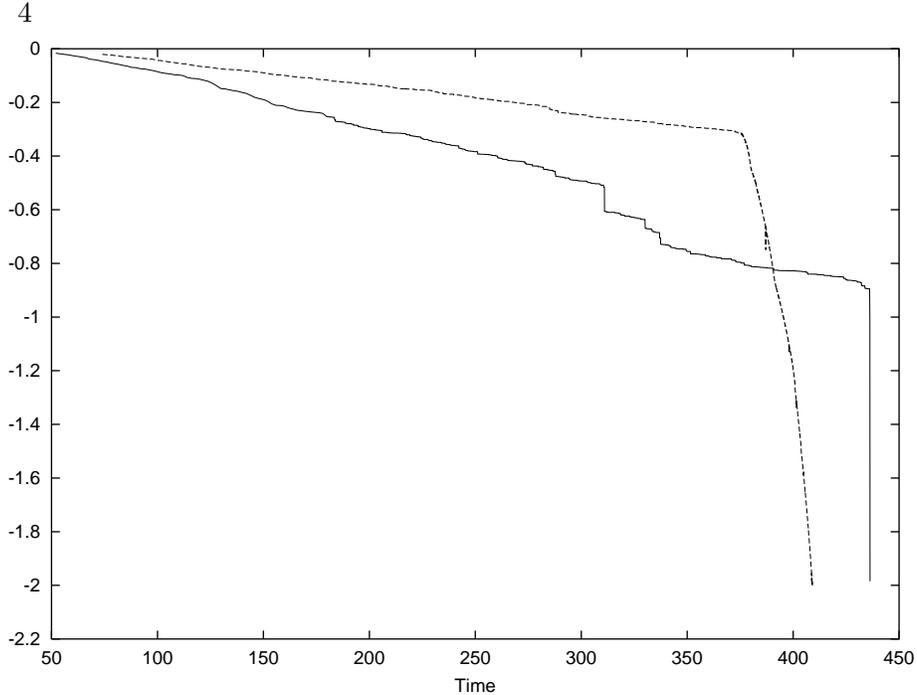}
\caption{\label{fig1}  Binding energy of BH binaries.
The upper curve (dashed line) is for $N = 2.4 \times 10^5$ and the lower
curve (solid line) for $N = 1.2 \times 10^5$.
Time is in scaled $N$-body units where one initial crossing time
is 2.8 or about 1\,Myr.
 }
\end{figure}

The increase of the binding energy, $E_{\rm BH} = - m_{\rm BH}^2 /2a$, is
illustrated in Fig. 1.
As a result of the scaling procedure for two subsystems, the initial total
energy is $E_{\rm tot} = -1.05$.
Not shown on the plot is the final value $E_{\rm BH} = -61$ with the
corresponding semi-major axis $a = 2.7 \times 10^{-7}$ for the smallest
system, compared to $r_{\rm h} \simeq 1$ initially.
Scale factors $r_{\rm h} = 4$\,pc and $\bar m = 1\,M_{\odot}$ were
chosen.
In other experiments we have demonstrated that a separation of three
Schwarzschild radii can be reached without numerical problems.
However, the end result of coalescence is ensured once the eccentricity
starts to decline significantly, in which case the present purpose is
achieved.

The strong binary evolution gives rise to the ejection of high-velocity
particles by the slingshot mechanism.
Although the effective mass ratio is about 1000 in the largest system,
these ejections still result in significant recoil velocities acquired
by the binary.
Hence the typical velocity of the central object exceeds the standard
equipartition value by a considerable factor, which has implications
for the so-called loss-cone effect.
However, it should be emphasized that the present results cannot be
scaled directly to systems with much larger mass ratios.

The eccentricity evolution of the two systems is shown in Fig. 2.
Although the initial eccentricities are relatively high, the trend is
for a gradual increase superimposed on fluctuations due to external
perturbations.
Two subsidiary maxima, $e_{\rm max} \simeq 0.998$ and 0.997, are first
reached in the smaller system, followed by temporary declines before the
final stage where coalescence sets in.
The eccentricity growth is more pronounced in the second model.
Note that $\dot e < 0$ during the final approach to coalescence.

The above examples should be considered as tests of the method rather
than giving definite results.
In this respect the outcome was highly successful, demonstrating that
the numerical scheme is both efficient and accurate.
Needless to say, the very large number of binary periods involved
($\sim 10^7$) represents a massive computational effort.
However, the binary BH problem is a fundamental one and its study by the
direct numerical approach is bound to be fruitful.

\begin{figure}[t]
\vspace{8.5cm}
\includegraphics{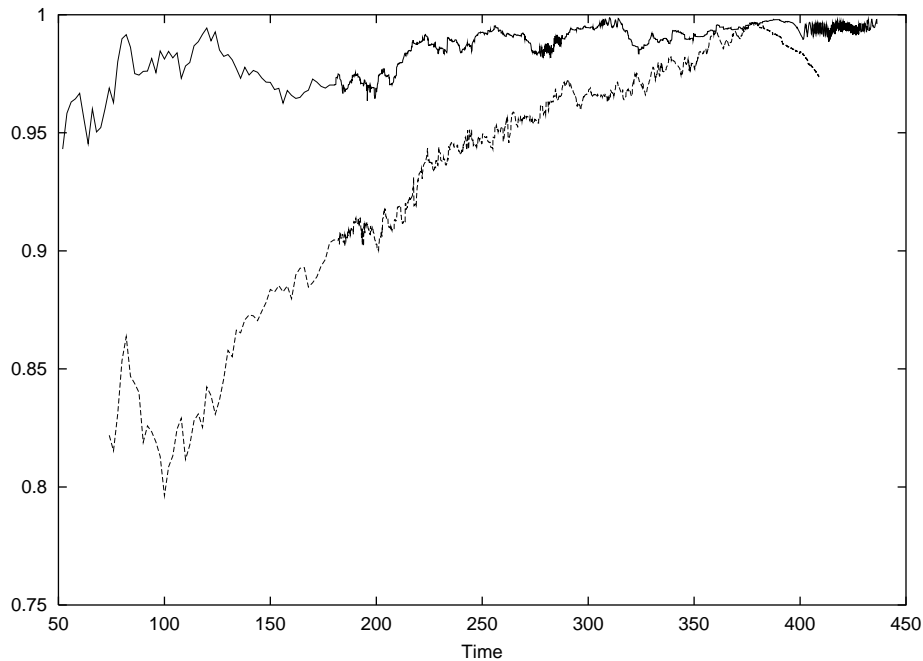}
\caption{\label{fig2}  Eccentricity evolution of BH binaries.
The upper curve is for $N = 1.2 \times 10^5$ and the lower curve
for $N = 2.4 \times 10^5$.
 }
\end{figure}

\end{article}
\end{document}